\documentclass[%
 reprint,
nolongbibliography,
 amsmath,amssymb,
 aps,
 prl
]{revtex4-2}

\usepackage{graphicx}
\usepackage{dcolumn}
\usepackage{bm}
\usepackage{hyperref}
\usepackage{natbib}

\begin{document}


\title{Thermodynamic Evidence of Tetracritical Topology in the
\texorpdfstring{$H$-$T$}{H-T} Phase Diagram of
\texorpdfstring{UTe$_2$}{UTe2} for
\texorpdfstring{$H \parallel b$}{H parallel b}}

\author{Michal Vali\v{s}ka}
 \email{michal.valiska@matfyz.cuni.cz}
 \affiliation{%
 Charles University, Faculty of Mathematics and Physics, Department of Condensed Matter Physics, Ke Karlovu 5, 121 16 Prague 2, Czech Republic}%
 
\author{Tetiana Haidamak}%
 \affiliation{%
 Charles University, Faculty of Mathematics and Physics, Department of Condensed Matter Physics, Ke Karlovu 5, 121 16 Prague 2, Czech Republic}%

\author{Andrej Cabala}%
 \affiliation{%
 Charles University, Faculty of Mathematics and Physics, Department of Condensed Matter Physics, Ke Karlovu 5, 121 16 Prague 2, Czech Republic}%

\author{Petr Proschek}%
 \affiliation{%
 Charles University, Faculty of Mathematics and Physics, Department of Condensed Matter Physics, Ke Karlovu 5, 121 16 Prague 2, Czech Republic}%

\author{Andreas Hauspurg}%
 \affiliation{Hochfeld-Magnetlabor Dresden (HLD-EMFL), Helmholtz-Zentrum Dresden-Rossendorf (HZDR), 01328 Dresden, Germany}%

\author{Sergei Zherlitsyn}%
 \affiliation{Hochfeld-Magnetlabor Dresden (HLD-EMFL), Helmholtz-Zentrum Dresden-Rossendorf (HZDR), 01328 Dresden, Germany}%

\author{Vladim\'{\i}r Sechovsk\'{y}}%
 \affiliation{%
 Charles University, Faculty of Mathematics and Physics, Department of Condensed Matter Physics, Ke Karlovu 5, 121 16 Prague 2, Czech Republic}%

\date{\today}

\begin{abstract}
We report ultrasound velocity measurements on an ultraclean UTe$_2$ single crystal with $T_c>2$~K for $H \parallel b$, performed up to 18~T and down to 0.33~K. The measurements provide the missing bulk thermodynamic evidence for an additional high-field phase boundary near $\mu_0H\sim14$--15~T. A distinct, non-hysteretic anomaly in the longitudinal $C_{33}$ mode, together with a weaker response in $C_{44}$, no resolvable anomaly in $C_{55}$, and a coincident kink in transverse magnetostriction, reveals a symmetry-selective coupling to lattice strain. This mode selectivity places constraints on the symmetry of the field-induced superconducting component. The phase line remains nearly field-constant near 14~T and meets three other phase boundaries at a tetracritical point near 13.5~T and 1.25~K. The results complete the local tetracritical topology of the $H$--$T$ phase diagram and support field-induced multicomponent superconductivity in UTe$_2$.
\end{abstract}

\maketitle


Since the discovery of superconductivity in UTe$_2$ by Ran \textit{et al.}, this heavy-fermion paramagnet has become one of the central platforms for the study of spin-triplet and potentially topological superconductivity \cite{Ran2019}. Early thermodynamic and transport studies rapidly established UTe$_2$ as an unconventional superconductor with a large and highly anisotropic upper critical field, strong-coupling behavior, and a striking sensitivity to magnetic field, pressure, and sample quality \cite{Aoki2019_UTe2,Braithwaite2019,Aoki_2022}. High-field and high-pressure measurements subsequently revealed a remarkably rich superconducting phase diagram whose detailed topology remains intensely debated \cite{Knafo2019a,Knebel2019,Knebel2020,Aoki2021,Knafo2021,Valiska2021,Sakai2023,Vasina2025PRL}.

The interpretation of the physical properties of UTe$_2$ is strongly affected by sample quality. Molten-salt-flux growth and related liquid-transport methods enabled ultraclean crystals with $T_c>2$~K \cite{Sakai2022,Aoki2024MSFLT,Wu2024PNAS}, while uranium deficiency and uranium-site defects were identified as key sources of suppressed or split transitions \cite{Haga2022,Svanidze2025}. The same ultraclean generation of crystals also enabled quantum-oscillation and quantum-interference studies of the quasi-two-dimensional Fermi surface \cite{Aoki2022dHvA,Eaton2024,Weinberger2024PRL}, and revealed the enhancement of superconductivity in intense magnetic fields \cite{Wu2025PNAS}. This issue is crucial for the high-field phase diagram for $H \parallel b$, where ultraclean crystals revealed additional anomalies near $\mu_0H \sim 14$--15~T in ac susceptibility and transport studies \cite{Sakai2023,Matsumura2026,Tokiwa2023PRB}. These anomalies were interpreted as an internal superconducting phase boundary, but a bulk thermodynamic signature has remained missing. A multicomponent spin-triplet scenario has also predicted a nearly field-constant internal line that meets the other three phase transition lines at a tetracritical point near 15~T \cite{Machida2024}.

This question has acquired additional interest through recent progress on the $p$-$T$ phase diagram. In a broader $H$-$p$-$T$ phase space, two locations have emerged where only three second-order phase boundaries appeared to meet, implying an incomplete and therefore thermodynamically problematic topology unless an additional phase line is present. Within a conventional Landau description of two distinct superconducting order parameters carrying the usual charge-$2e$ gauge transformation, three continuous phase boundaries cannot generically terminate at a single point: a direct boundary between two single-component ordered phases is generically first order, whereas a tetracritical topology requires a coexistence phase and four continuous boundaries \cite{SigristUeda1991,Yip1991,Kamat2026}. A recent alternative proposal has argued that a triple point formed by three continuous transition lines is not forbidden on purely thermodynamic grounds if superconducting order parameters with different gauge-transformation laws are allowed \cite{Melnikovsky2026}. This work does not rely on excluding this possibility in principle, but on directly resolving an additional bulk thermodynamic phase boundary.

One of these cases, in the $p$-$T$ plane, has now been resolved by recent ultrasound measurements that uncovered the missing thermodynamic boundary and established a tetracritical point (TetCP) \cite{Kamat2026}. At the same time, the high-field and high-pressure superconducting phases were shown to be continuously connected \cite{Vasina2025PRL}, and the corresponding $H$-$p$-$T$ construction implies that the pressure-induced TetCP evolves into a tetracritical line extending toward ambient pressure and the high-field superconducting regime \cite{Kamat2026}. Within this picture, the long-suspected anomaly near 14--15~T for $H \parallel b$ is naturally expected to represent the missing internal phase boundary whose endpoint at ambient pressure forms the high-field TetCP. However, this ambient-pressure high-field continuation has so far been inferred from the pressure-dependent phase diagram rather than established directly by a bulk thermodynamic measurement \cite{Kamat2026}.

Here we address this issue using pulse-echo ultrasound measurements of the elastic constants $C_{ij}$ on an ultraclean UTe$_2$ crystal with $T_c > 2$~K for magnetic field applied along the $b$ axis. The experimental geometry, pulse-echo analysis, sample preparation, and field-sweep protocol are described in the Supplemental Material \cite{SupplementalMaterial}. We resolve a distinct non-hysteretic anomaly in $C_{33}(H)$ within the superconducting state that traces a nearly field-constant line near 14~T between 0.33 and 1.1~K. We show that this anomaly constitutes bulk thermodynamic evidence for the fourth phase line in the $H$-$T$ phase diagram. The line terminates at a TetCP near 13.5~T and 1.25~K, consistent with tetracritical topology proposed theoretically \cite{Machida2024} and with the tetracritical-line scenario inferred from pressure studies \cite{Kamat2026}. Our results place the high-field $H$-$T$ phase diagram of UTe$_2$ for $H \parallel b$ on firm thermodynamic footing and provide further support for multicomponent superconductivity in the ultraclean limit.

The phase boundary sought in this work is examined through its coupling to lattice strain $\varepsilon_{ij}$. Elastic constants and magnetostriction are particularly suitable for this purpose, because both are thermodynamic probes of the same strain-dependent free-energy landscape \cite{Luthi2005,SigristUeda1991}. Elastic constants are determined by the strain dependence of the free-energy density $f$ as
\begin{equation}
C_{ij}
=
\left(\frac{\partial^2 f}{\partial \varepsilon_i \partial \varepsilon_j}\right)_{T,H}.
\label{eq:Cij_free_energy}
\end{equation}

An anomaly in $C_{ij}$ therefore reflects a change in the local stiffness of the strain-dependent free-energy landscape and represents a bulk thermodynamic signature of a phase transition or phase boundary \cite{Luthi2005,SigristUeda1991}.

Magnetostriction provides a complementary static probe by tracking the field evolution of the equilibrium strain.

The geometry was chosen to maximize sensitivity to the high-field boundary. Superconductivity in UTe$_2$ couples strongly to $c$-axis strain but only weakly to $b$-axis strain \cite{Theuss2023}, motivating our focus on $C_{33}$ for $H \parallel b$. The complementary transverse magnetostriction $\Delta l_c/l_c$ probes the same $c$-axis strain channel, whereas earlier high-field magnetostriction work addressed the longitudinal response $\Delta l_b/l_b$ \cite{Rosuel2023}.

\begin{figure*}[t]
    \centering
    \includegraphics[width=\textwidth]{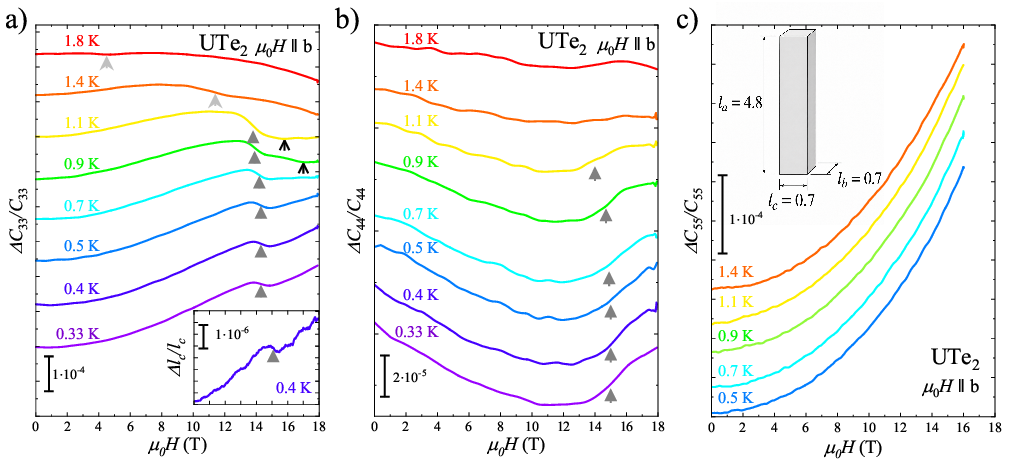}
    \caption{
    Field dependence of the elastic response for $H \parallel b$.
(a) Longitudinal mode $C_{33}(H)$ measured with $\mathbf{q}\parallel c$ and $\mathbf{u}\parallel c$, showing a pronounced softening anomaly near 14~T. The inset shows a selected field window of the transverse magnetostriction $\Delta l_c/l_c$ for $H \parallel b$.
(b) Transverse mode $C_{44}(H)$ measured with $\mathbf{q}\parallel c$ and $\mathbf{u}\parallel b$, where a weaker hardening anomaly is observed in the same field range.
(c) Transverse mode $C_{55}(H)$ measured with $\mathbf{q}\parallel c$ and $\mathbf{u}\parallel a$, where no corresponding anomaly is resolved within the experimental resolution; the inset shows the measured sample and its dimensions in millimeters.
Arrows with different styles mark the anomalies used to construct the phase diagram in Fig.~\ref{fig:phase_diagram}.
All elastic-constant traces shown are increasing-field sweeps.
Direct comparisons of increasing- and decreasing-field $C_{33}(H)$
sweeps across the internal boundary are provided in the Supplemental
Material \cite{SupplementalMaterial} and show no measurable hysteresis
within the experimental resolution.
Vertical scales are indicated by the scale bars in each panel.
    }
    \label{fig:field_sweeps}
\end{figure*}

Figure~\ref{fig:field_sweeps}(a) displays the field dependence of the longitudinal mode $C_{33}$ at selected temperatures. A clear softening anomaly, marked by arrows, is resolved inside the superconducting state with a field position that remains close to 14~T over the measured temperature range. Since $C_{33}$ is a bulk elastic constant, this anomaly provides the missing thermodynamic signature of the high-field feature previously inferred only from ac susceptibility and transport measurements \cite{Sakai2023,Tokiwa2023PRB,Matsumura2026}, and supports its identification as an internal superconducting phase boundary. The anomaly is visible from the lowest measured temperature up to 1.1~K, and its amplitude increases substantially on approaching the multicritical region where the line terminates at the TetCP. For the 0.9 and 1.1~K isotherms, an additional kink is observed at higher field, consistent with the transition into the upper high-field superconducting phase \cite{Knebel2019,Knafo2021,Wu2025PNAS}. At temperatures above the TetCP, the internal-boundary anomaly associated with the nearly field-constant line is no longer resolved. Instead, a separate anomaly shifts to lower fields and is associated with the boundary between the low-field superconducting state and the normal state.

The $C_{33}(H)$ curves used to track the internal boundary were
acquired on increasing field. At all temperatures for which both sweep
directions were recorded, the corresponding increasing- and
decreasing-field curves overlap within the experimental resolution, as
shown by direct comparison in the Supplemental Material
\cite{SupplementalMaterial}. The absence of measurable hysteresis strongly disfavors an interpretation in terms of irreversible vortex pinning or depinning, although equilibrium vortex-matter transitions may themselves produce reversible thermodynamic signatures once pinning-induced irreversibility is removed \cite{Avraham2001}.

The enhancement of the $C_{33}$ anomaly on approaching the TetCP is consistent with the sensitivity of elastic constants to critical and multicritical fluctuations in correlated uranium systems \cite{Valiska2024PRM,Yoshizawa2023,Haidamak2022}.

A weaker anomaly is also observed in the transverse mode $C_{44}$ [Fig.~\ref{fig:field_sweeps}(b)]. It has the opposite sign, appearing as a hardening, and its field position tracks the $C_{33}$ anomaly, indicating that both modes detect the same boundary. By contrast, no corresponding anomaly is resolved in the transverse $C_{55}$ mode within our experimental resolution [Fig.~\ref{fig:field_sweeps}(c)]. This mode selectivity mirrors the pressure-induced tetracritical case, where the newly identified internal boundary under pressure was prominent in the $C_{33}$ elastic response but not comparably visible in $C_{55}$ \cite{Kamat2026}.

The inset of Fig.~\ref{fig:field_sweeps}(a) shows the corresponding transverse magnetostriction $\Delta l_c/l_c$ for $H \parallel b$. A kink is observed at essentially the same field as the anomaly in $C_{33}$, confirming that this boundary affects not only the dynamic elastic stiffness but also the static equilibrium lattice distortion. The coincidence of the two anomalies provides a complementary static thermodynamic signature of the same high-field feature.

The mode selectivity also has symmetry implications. In orthorhombic $D_{2h}$ symmetry, the shear modes probe different strain channels: $C_{44}$ couples to $\varepsilon_{bc}$, $C_{55}$ to $\varepsilon_{ac}$, and $C_{66}$ to $\varepsilon_{ab}$ \cite{SigristUeda1991,Theuss2023}. Recent ultrasound work argued that the low-field superconducting state is single-component, with an odd-parity $B_{2u}$ state as the most compatible candidate, based on the absence of shear-modulus discontinuities at $T_c$ \cite{Theuss2023}. Because a magnetic field along $b$ lowers the point-group symmetry from $D_{2h}$ to $C_{2h}$, the $D_{2h}$ representations used above should be understood as parent-state labels, and components that belong to the same reduced-symmetry representation may in principle mix in field \cite{Kamat2026}.

Taking this proposed $B_{2u}$ assignment as a working hypothesis and using the axis convention $x\parallel a$, $y\parallel b$, and $z\parallel c$, the longitudinal $C_{33}$ mode probes the fully symmetric $A_g$ strain channel and can couple to the strain dependence of any superconducting component; by itself, it therefore does not identify the representation of a field-induced component. By contrast, the shear strains transform as $\varepsilon_{bc}\sim B_{3g}$, $\varepsilon_{ac}\sim B_{2g}$, and $\varepsilon_{ab}\sim B_{1g}$, corresponding to $C_{44}$, $C_{55}$, and $C_{66}$, respectively. Assuming a predominantly $B_{2u}$ low-field state and the lowest-order bilinear strain coupling between two superconducting components, a shear strain $\varepsilon_{\Gamma}$ can couple to the bilinear combination of the two order parameters only when $\Gamma$ matches their product representation. In this convention, a secondary $B_{1u}$ component couples to $\varepsilon_{bc}$ and hence to $C_{44}$, a secondary $A_u$ component couples to $\varepsilon_{ac}$ and hence to $C_{55}$, and a secondary $B_{3u}$ component would couple to $\varepsilon_{ab}$ and hence to $C_{66}$. The observed $C_{44}$ anomaly together with the absence of a resolvable $C_{55}$ response is therefore consistent with coupling through the $B_{3g}$ shear channel and makes dominant coupling through the $B_{2g}$ channel expected for an $A_u$ secondary component less likely. A $B_{3u}$ component would instead couple through the unmeasured $C_{66}$ channel and is therefore not excluded by the present data. These conclusions serve as symmetry constraints rather than a unique assignment. They assume that the low-field state is predominantly $B_{2u}$, that the zero-field $D_{2h}$ classification remains a useful parent-state basis for $H\parallel b$, and that the lowest-order bilinear shear coupling dominates. Alternative mixed low-field states, including the chiral $B_{3u}+iA_u$ state proposed from anisotropic penetration-depth measurements \cite{Ishihara2023}, cannot be excluded, nor can differences among the individual strain-coupling constants \cite{SigristUeda1991,Theuss2023}. Within these limitations, the mode selectivity nevertheless provides thermodynamic constraints on the field-induced order-parameter structure and is compatible with multicomponent scenarios for the tetracritical region \cite{Machida2024,Kappler1995,Wu2025PRL,Kamat2026}.

\begin{figure}[t]
    \centering
    \includegraphics[width=\columnwidth]{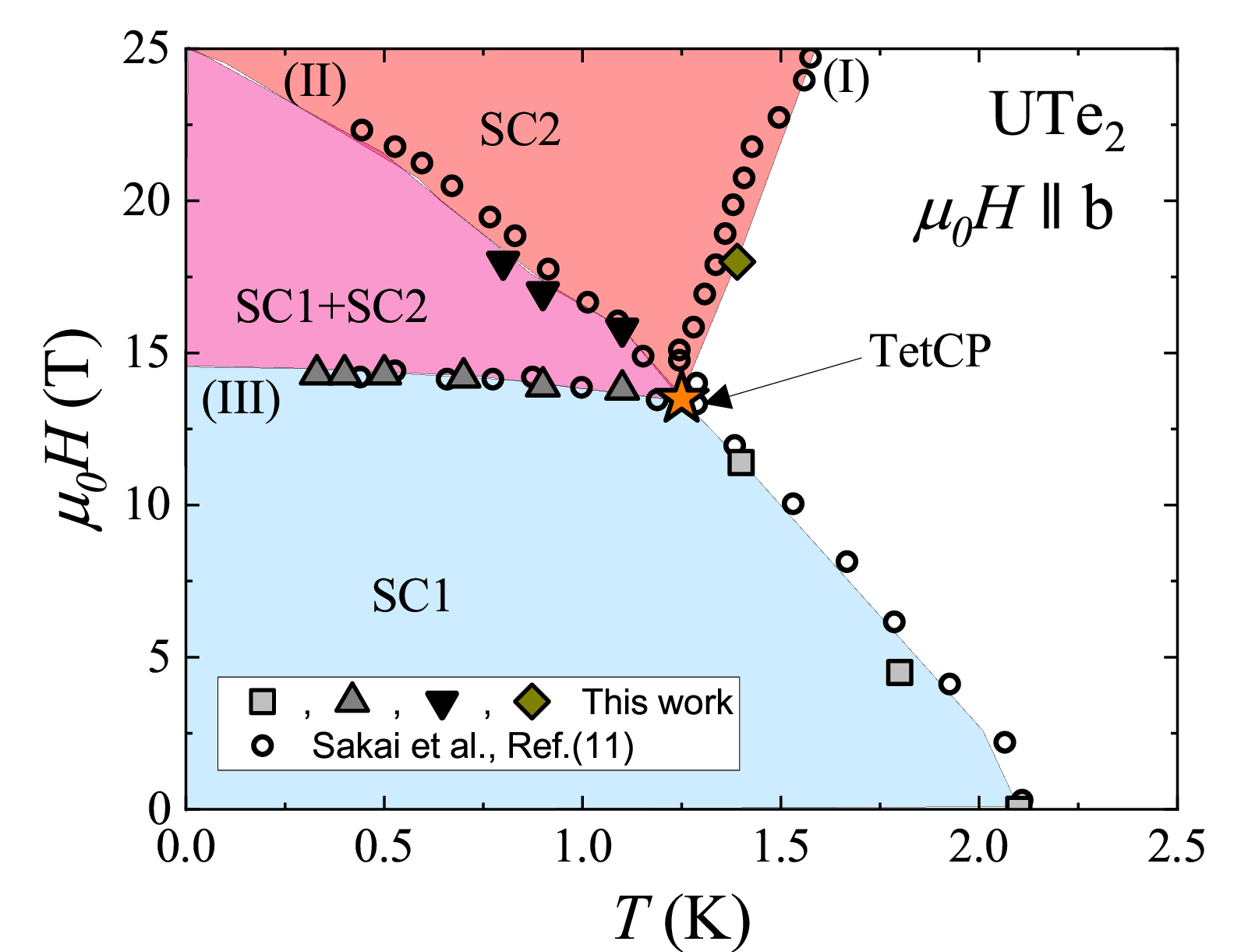}
    \caption{
$H$-$T$ phase diagram of UTe$_2$ for $H \parallel b$.
Filled symbols denote phase boundaries determined in the present work; open symbols and the notation (I)-(III) for the high-field boundaries follow Sakai \textit{et al.} \cite{Sakai2023}.
The fourth boundary meeting the tetracritical point is the low-field $H_{c2}^{\rm SC1}$ line.
The thermodynamically identified internal boundary, line (III), terminates at the TetCP near 13.5~T and 1.25~K, completing the local phase-diagram topology.
The superconducting regions SC1, SC1+SC2, and SC2 are indicated.
    }
    \label{fig:phase_diagram}
\end{figure}

Complementary temperature sweeps of $C_{33}$, $C_{44}$, and $C_{55}$ at selected fields are presented in the Supplemental Material \cite{SupplementalMaterial}. The corresponding anomalies provide additional transition points used to construct the phase diagram in Fig.~\ref{fig:phase_diagram}.

The anomaly positions extracted from the thermodynamic data are summarized in Fig.~\ref{fig:phase_diagram}. Following the notation of the pressure-dependent phase diagram \cite{Vasina2025PRL} and its thermodynamic and Ginzburg--Landau construction \cite{Kamat2026}, we denote the low-field superconducting state as SC1 and the upper high-field state as SC2. The intermediate region beyond the internal boundary is then identified as SC1+SC2, because, within this construction, the boundary is continuously connected to the SC2 phase boundary and separates the pure SC1 state from a region in which the SC1 and SC2 order parameters coexist. The three high-field boundaries are labeled (I), (II), and (III) following Sakai \textit{et al.} \cite{Sakai2023}, and separate SC2 from the normal state, SC2 from SC1+SC2, and SC1+SC2 from SC1, respectively. The fourth boundary meeting the TetCP is the low-field $H_{c2}^{\rm SC1}$ boundary. The thermodynamic anomaly coincides with line (III) and provides bulk evidence that this boundary terminates at the TetCP, where all four phase boundaries meet. Thus, the elastic-constant anomaly supplies the missing bulk thermodynamic evidence needed to complete the local $H\parallel b$ phase-diagram topology and supports field-induced multicomponent superconductivity. This topology is consistent with the tetracritical-line scenario inferred from the pressure-dependent phase diagram \cite{Machida2024,Kamat2026}, and echoes the role of sound velocity in sensing tetracritical topology in UPt$_3$ \cite{Boukhny1994}.

The present ambient-pressure endpoint near 13.5~T and 1.25~K is
semiquantitatively consistent with the broader $H$-$p$-$T$ construction
of Kamat \textit{et al.}, in which the pressure-induced TetCP evolves
toward the ambient-pressure high-field regime \cite{Kamat2026}.
The moderately higher field and temperature scales found here may reflect
the higher $T_c$ and ultraclean character of the present crystal, given
the pronounced sample dependence of the superconducting phase diagram of
UTe$_2$.
Importantly, the corresponding SC1--SC1+SC2 boundary is observed under
pressure already at zero magnetic field, where a vortex-lattice origin is
absent, and is subsequently tracked to finite field. The assignment of
the present anomaly therefore rests not only on its reversibility, but
also on its continuous connection to this independently established
boundary. A vortex interpretation would require a fortuitous coincidence
with the continuation of the same zero-field pressure-induced phase line,
making the multicomponent superconducting interpretation substantially
more natural.

The local geometry in the neighborhood of the TetCP also helps explain why line (III) has been difficult to detect calorimetrically. The corresponding local slopes $dH/dT$ for $H_{c2}^{\rm SC1}$, (I), (II), and (III) are approximately $-14$, $+34$, $-12$, and $-0.5$~T/K, respectively. Line (III) therefore has a nearly field-constant position near 14~T, with a much smaller $|dH/dT|$ than the other three boundaries.

For a field-tuned second-order boundary $H^\ast(T)$, the Ehrenfest relation gives, up to conventional prefactors,
\begin{equation}
\Delta\!\left(\frac{C_p}{T}\right)
\propto
\Delta\chi
\left(\frac{dH^\ast}{dT}\right)^2 ,
\label{eq:Ehrenfest_HT}
\end{equation}
where $\Delta\chi$ is the discontinuity in the magnetic susceptibility across the boundary. Thus, the calorimetric anomaly is quadratically suppressed when $dH^\ast/dT$ is small \cite{Yip1991,Adenwalla1992,Boukhny1994}. Since the slope of line (III) is more than one order of magnitude smaller than those of the other three boundaries, the expected specific-heat anomaly is reduced by roughly three to four orders of magnitude, for comparable $\Delta\chi$. This naturally explains why the internal 14--15~T phase boundary may remain essentially unresolved in specific heat, even if it is a genuine thermodynamic transition \cite{Machida2024,Sakai2023,Vasina2025PRL,Vasina2026Calorimetry}. 

The elastic response provides a complementary route because it is governed by a different derivative of the same field-tuned boundary. For $H^\ast(T,\varepsilon)$, the corresponding Ehrenfest relation may be written, again up to conventional prefactors, as
\begin{equation}
\Delta C_{ij}
\propto
-\Delta\chi
\left(\frac{\partial H^\ast}{\partial \varepsilon_i}\right)_{T}
\left(\frac{\partial H^\ast}{\partial \varepsilon_j}\right)_{T}.
\label{eq:Ehrenfest_elastic_field}
\end{equation}
Equations~(\ref{eq:Ehrenfest_HT}) and (\ref{eq:Ehrenfest_elastic_field}) therefore highlight why the same transition can be negligible in $C_p$ but prominent in elasticity: the calorimetric anomaly is suppressed by the small slope $dH^\ast/dT$, whereas the elastic anomaly remains large if the critical field is strongly strain dependent. The clear $C_{33}$ anomaly and the corresponding feature in $\Delta l_c/l_c$ indicate precisely such a strong coupling to $c$-axis strain, making ultrasound and magnetostriction particularly sensitive probes of this otherwise elusive phase line \cite{Luthi2005,SigristUeda1991}.

This interpretation is also supported by recent quantitative calorimetry under pressure, where the $C_p$ anomaly associated with SC2 was found to extrapolate to a nearly vanishing jump at the point where the SC2 line meets the SC1 boundary \cite{Vasina2026Calorimetry}, whereas ultrasound measurements resolved the corresponding internal thermodynamic boundary and established the pressure-induced TetCP \cite{Kamat2026}. This reinforces the central point that an internal boundary may be weak or nearly invisible in specific heat while remaining clearly detectable through its coupling to lattice strain.

Taken together, the present results provide bulk thermodynamic evidence for an internal boundary near 14--15~T for $H \parallel b$. In conjunction with the independently established pressure-dependent phase diagram, this supports identifying line (III) as the fourth boundary of the ambient-pressure TetCP. The mode-selective elastic response constrains the symmetry of the field-induced order-parameter structure, while the coincident magnetostriction anomaly confirms its coupling to the equilibrium lattice strain. Within this interpretation, crossing line (III) corresponds to entering a mixed SC1+SC2 phase, linking the ambient-pressure high-field phase diagram to the pressure-induced multicomponent regime \cite{Theuss2023,Vasina2025PRL,Kamat2026}.

\noindent\textit{Acknowledgments.---}
The UTe$_2$ crystals were grown and characterized at MGML (\url{https://mgml.eu/}), supported within the program of Czech Research Infrastructures (project no. LM2023065). We acknowledge support from the Dresden High Magnetic Field Laboratory (HLD) at Helmholtz-Zentrum Dresden-Rossendorf (HZDR), a member of the European Magnetic Field Laboratory (EMFL). This work was supported by the Czech Science Foundation (GA\v{C}R) under the Junior Star Grant No. 26-21795M (STiUS).

\bibliography{references}

@article{Knafo2021,
  author  = {Knafo, W. and Nardone, M. and Vališka, M. and Zitouni, A. and Lapertot, G. and Aoki, D. and Knebel, G. and Braithwaite, D.},
  title   = {Comparison of two superconducting phases induced by a magnetic field in {UTe$_2$}},
  journal = {Communications Physics},
  volume  = {4},
  number  = {1},
  pages   = {40},
  year    = {2021},
  doi     = {10.1038/s42005-021-00545-z}
}

@article{Aoki_2022,
  author  = {Aoki, D. and Brison, J.-P. and Flouquet, J. and Ishida, K. and Knebel, G. and Tokunaga, Y. and Yanase, Y.},
  title   = {Unconventional superconductivity in {UTe$_2$}},
  journal = {Journal of Physics: Condensed Matter},
  volume  = {34},
  number  = {24},
  pages   = {243002},
  year    = {2022},
  doi     = {10.1088/1361-648X/ac5863}
}

@article{Knafo2019a,
  author  = {Knafo, William and Vališka, Michal and Braithwaite, Daniel and Lapertot, Gérard and Knebel, Georg and Pourret, Alexandre and Brison, Jean-Pascal and Flouquet, Jacques and Aoki, Dai},
  title   = {Magnetic-Field-Induced Phenomena in the Paramagnetic Superconductor {UTe$_2$}},
  journal = {Journal of the Physical Society of Japan},
  volume  = {88},
  number  = {6},
  pages   = {063705},
  year    = {2019},
  doi     = {10.7566/JPSJ.88.063705}
}

@book{Luthi2005,
  author    = {Lüthi, B.},
  title     = {Physical Acoustics in the Solid State},
  publisher = {Springer},
  address   = {Berlin and Heidelberg},
  year      = {2005}
}

@article{Yoshizawa2023,
  author  = {Yoshizawa, Masahito and Shimizu, Yusei and Nakanishi, Yoshiki and Homma, Yoshiya and Nakamura, Ai and Honda, Fuminori and Aoki, Dai},
  title   = {Emergence of Elastic Softening Featuring Ultra-Slow Dynamics Around Magnetic Critical Endpoint in {UCoAl}},
  journal = {Journal of the Physical Society of Japan},
  volume  = {92},
  number  = {10},
  pages   = {104603},
  year    = {2023},
  doi     = {10.7566/JPSJ.92.104603}
}

@article{Sakai2022,
  author  = {Sakai, H. and Opletal, P. and Tokiwa, Y. and Yamamoto, E. and Tokunaga, Y. and Kambe, S. and Haga, Y.},
  title   = {Single crystal growth of superconducting {UTe$_2$} by molten salt flux method},
  journal = {Physical Review Materials},
  volume  = {6},
  number  = {7},
  pages   = {073401},
  year    = {2022},
  doi     = {10.1103/PhysRevMaterials.6.073401}
}

@article{Eaton2024,
  author  = {Eaton, A. G. and Weinberger, T. I. and Popiel, N. J. M. and Wu, Z. and Hickey, A. J. and Cabala, A. and Pospíšil, J. and Prokleška, J. and Haidamak, T. and Bastien, G. and Opletal, P. and Sakai, H. and Haga, Y. and Nowell, R. and Benjamin, S. M. and Sechovský, V. and Lonzarich, G. G. and Grosche, F. M. and Vališka, M.},
  title   = {Quasi-2D Fermi surface in the anomalous superconductor {UTe$_2$}},
  journal = {Nature Communications},
  volume  = {15},
  number  = {1},
  pages   = {223},
  year    = {2024},
  doi     = {10.1038/s41467-023-44110-4}
}

@article{Haidamak2022,
  author  = {Haidamak, T. N. and Valenta, J. and Prchal, J. and Vališka, M. and Pospíšil, J. and Sechovský, V. and Prokleška, J. and Zvyagin, A. A. and Honda, F.},
  title   = {Tricritical fluctuations and elastic properties of the {Ising} antiferromagnet {UIrSi$_3$}},
  journal = {Physical Review B},
  volume  = {105},
  number  = {14},
  pages   = {144428},
  year    = {2022},
  doi     = {10.1103/PhysRevB.105.144428}
}

@article{Ran2019,
  author  = {Ran, Sheng and Eckberg, Chris and Ding, Qing-Ping and Furukawa, Yuji and Metz, Tristin and Saha, Shanta R. and Liu, I-Lin and Zic, Mark and Kim, Hyunsoo and Paglione, Johnpierre and Butch, Nicholas P.},
  title   = {Nearly ferromagnetic spin-triplet superconductivity},
  journal = {Science},
  volume  = {365},
  number  = {6454},
  pages   = {684--687},
  year    = {2019},
  doi     = {10.1126/science.aav8645}
}

@article{Theuss2023,
  author  = {Theuss, Florian and Shragai, Avi and Grissonnanche, Gaël and Hayes, Ian M. and Saha, Shanta R. and Eo, Yun Suk and Suarez, Alonso and Shishidou, Tatsuya and Butch, Nicholas P. and Paglione, Johnpierre and Ramshaw, B. J.},
  title   = {Single-component superconductivity in {UTe$_2$} at ambient pressure},
  journal = {Nature Physics},
  volume  = {20},
  pages   = {1124},
  year    = {2024},
  doi     = {10.1038/s41567-024-02493-1}
}

@article{Weinberger2024PRL,
  author  = {Weinberger, T. I. and Wu, Z. and Graf, D. E. and Skourski, Y. and Cabala, A. and Pospíšil, J. and Prokleška, J. and Haidamak, T. and Bastien, G. and Sechovský, V. and Lonzarich, G. G. and Vališka, M. and Grosche, F. M. and Eaton, A. G.},
  title   = {Quantum Interference between Quasi-2D Fermi Surface Sheets in {UTe$_2$}},
  journal = {Physical Review Letters},
  volume  = {132},
  number  = {26},
  pages   = {266503},
  year    = {2024},
  doi     = {10.1103/PhysRevLett.132.266503}
}

@article{Wu2024PNAS,
  author  = {Wu, Z. and Weinberger, T. I. and Chen, J. and Cabala, A. and Chichinadze, D. V. and Shaffer, D. and Pospíšil, J. and Prokleška, J. and Haidamak, T. and Bastien, G. and Sechovský, V. and Hickey, A. J. and Mancera-Ugarte, M. J. and Benjamin, S. and Graf, D. E. and Skourski, Y. and Lonzarich, G. G. and Vališka, M. and Grosche, F. M. and Eaton, A. G.},
  title   = {Enhanced triplet superconductivity in next-generation ultraclean {UTe$_2$}},
  journal = {Proc. Natl. Acad. Sci. U.S.A.},
  volume  = {121},
  number  = {37},
  pages   = {e2403067121},
  year    = {2024},
  doi     = {10.1073/pnas.2403067121}
}

@article{Aoki2019_UTe2,
  author  = {Aoki, Dai and Nakamura, Ai and Honda, Fuminori and Li, Dexin and Homma, Yoshiya and Shimizu, Yusei and Sato, Yoshiki J. and Knebel, Georg and Brison, Jean-Pascal and Pourret, Alexandre and Braithwaite, Daniel and Lapertot, Gérard and Niu, Qun and Vališka, Michal and Harima, Hisatomo and Flouquet, Jacques},
  title   = {Unconventional Superconductivity in Heavy Fermion {UTe$_2$}},
  journal = {Journal of the Physical Society of Japan},
  volume  = {88},
  number  = {4},
  pages   = {043702},
  year    = {2019},
  doi     = {10.7566/JPSJ.88.043702}
}

@article{Braithwaite2019,
  author  = {Braithwaite, D. and Vališka, M. and Knebel, G. and Lapertot, G. and Brison, J.-P. and Pourret, A. and Zhitomirsky, M. E. and Flouquet, J. and Honda, F. and Aoki, D.},
  title   = {Multiple superconducting phases in a nearly ferromagnetic system},
  journal = {Communications Physics},
  volume  = {2},
  pages   = {147},
  year    = {2019},
  doi     = {10.1038/s42005-019-0248-z}
}

@article{Haga2022,
  author  = {Haga, Yoshinori and Opletal, Petr and Tokiwa, Yoshifumi and Yamamoto, Etsuji and Tokunaga, Yo and Kambe, Shinsaku and Sakai, Hironori},
  title   = {Effect of uranium deficiency on normal and superconducting properties in unconventional superconductor {UTe$_2$}},
  journal = {Journal of Physics: Condensed Matter},
  volume  = {34},
  pages   = {175601},
  year    = {2022},
  doi     = {10.1088/1361-648X/ac5201}
}

@article{Sakai2023,
  author  = {Sakai, H. and Tokiwa, Y. and Opletal, P. and Kimata, M. and Awaji, S. and Sasaki, T. and Aoki, D. and Kambe, S. and Tokunaga, Y. and Haga, Y.},
  title   = {Field Induced Multiple Superconducting Phases in {UTe$_2$} along Hard Magnetic Axis},
  journal = {Physical Review Letters},
  volume  = {130},
  number  = {19},
  pages   = {196002},
  year    = {2023},
  doi     = {10.1103/PhysRevLett.130.196002}
}

@article{Aoki2024MSFLT,
  author  = {Aoki, Dai},
  title   = {Molten Salt Flux Liquid Transport Method for Ultra Clean Single Crystals {UTe$_2$}},
  journal = {Journal of the Physical Society of Japan},
  volume  = {93},
  number  = {4},
  pages   = {043703},
  year    = {2024},
  doi     = {10.7566/JPSJ.93.043703}
}

@article{Vasina2025PRL,
  author  = {Vasina, T. and Aoki, D. and Miyake, A. and Seyfarth, G. and Pourret, A. and Marcenat, C. and Amano Patino, M. and Lapertot, G. and Flouquet, J. and Brison, J.-P. and Braithwaite, D. and Knebel, G.},
  title   = {Connecting High-Field and High-Pressure Superconductivity in {UTe$_2$}},
  journal = {Physical Review Letters},
  volume  = {134},
  number  = {9},
  pages   = {096501},
  year    = {2025},
  doi     = {10.1103/PhysRevLett.134.096501}
}

@article{Svanidze2025,
  author  = {Svanidze, Eteri and Leithe-Jasper, Andreas and Schmidt, Marcus and Zaremba, Nazar and Krnel, Mitja and Prots, Yurii and Burkhardt, Ulrich and K{\"o}nig, Markus and Ramlau, Reiner and Goodge, Berit H. and Grin, Yuri},
  title   = {Chemical Facets of Superconductivity in {UTe$_2$}},
  journal = {Journal of the American Chemical Society},
  volume  = {147},
  number  = {39},
  pages   = {35809--35817},
  year    = {2025},
  doi     = {10.1021/jacs.5c12203}
}

@article{Matsumura2026,
  author  = {Matsumura, Hiroki and Takahashi, Yuki and Matsubayashi, Riku and Kinjo, Katsuki and Kitagawa, Shunsaku and Ishida, Kenji and Tokunaga, Yo and Sakai, Hironori and Kambe, Shinsaku and Kimata, Motoi and Nakamura, Ai and Shimizu, Yusei and Homma, Yoshiya and Li, Dexin and Honda, Fuminori and Miyake, Atsushi and Aoki, Dai and Furukawa, Tetsuya and Sasaki, Takahiro},
  title   = {Intimate relationship between spin configuration in the triplet pair and superconductivity in {UTe$_2$}},
  journal = {Physical Review B},
  volume  = {113},
  pages   = {094506},
  year    = {2026},
  doi     = {10.1103/xr3p-cxbt}
}

@misc{Kamat2026,
  author        = {Kamat, Sahas and Dans, Jared and Saha, Shanta and Kokovin, Artem D. and Paglione, Johnpierre and Schmalian, Jörg and Ramshaw, B. J.},
  year          = {2026},
  archivePrefix = {arXiv},
  eprint        = {2603.17905},
  primaryClass  = {cond-mat.supr-con},
  doi           = {10.48550/arXiv.2603.17905}
}

@article{Machida2024,
  author  = {Machida, Kazushige},
  title   = {Theoretical Studies on Off-Axis Phase Diagrams and Knight Shifts in {UTe$_2$}: Tetra-Critical Point, d-Vector Rotation, and Multiple Phases},
  journal = {Journal of Low Temperature Physics},
  volume  = {216},
  pages   = {746--788},
  year    = {2024},
  doi     = {10.1007/s10909-024-03181-3}
}

@article{Rosuel2023,
  author  = {Rosuel, A. and Marcenat, C. and Knebel, G. and Klein, T. and Pourret, A. and Marquardt, N. and Niu, Q. and Rousseau, S. and Demuer, A. and Seyfarth, G. and Lapertot, G. and Aoki, D. and Braithwaite, D. and Flouquet, J. and Brison, J. P.},
  title   = {Field-Induced Tuning of the Pairing State in a Superconductor},
  journal = {Physical Review X},
  volume  = {13},
  pages   = {011022},
  year    = {2023},
  doi     = {10.1103/PhysRevX.13.011022}
}

@article{Yip1991,
  author  = {Yip, S. K. and Li, T. and Kumar, P.},
  title   = {Thermodynamic considerations and the phase diagram of superconducting {UPt}$_3$},
  journal = {Physical Review B},
  volume  = {43},
  number  = {4},
  pages   = {2742--2747},
  year    = {1991},
  doi     = {10.1103/PhysRevB.43.2742}
}

@article{Adenwalla1992,
  author  = {Adenwalla, S. and Ketterson, J. B. and Yip, S. K. and Lin, S. W. and Levy, M. and Sarma, Bimal K.},
  title   = {Thermodynamics of the {UPt}$_3$ superconducting phase diagram},
  journal = {Physical Review B},
  volume  = {46},
  number  = {14},
  pages   = {9070--9082},
  year    = {1992},
  doi     = {10.1103/PhysRevB.46.9070}
}

@article{SigristUeda1991,
  author  = {Sigrist, M. and Ueda, K.},
  title   = {Phenomenological theory of unconventional superconductivity},
  journal = {Reviews of Modern Physics},
  volume  = {63},
  number  = {2},
  pages   = {239--311},
  year    = {1991},
  doi     = {10.1103/RevModPhys.63.239}
}

@article{Boukhny1994,
  author = {M. Boukhny and G. L. Bullock and B. S. Shivaram and D. G. Hinks},
  title = {Tetracritical Points and the Superconducting Phases of {UPt}$_3$: Uniaxial Pressure Effects},
  journal = {Physical Review Letters},
  volume = {73},
  number = {12},
  pages = {1707--1710},
  year = {1994},
  doi = {10.1103/PhysRevLett.73.1707},
  url = {https://doi.org/10.1103/PhysRevLett.73.1707}
}

@article{Kappler1995,
  author = {C. Kappler and M. B. Walker and J. Luettmer-Strathmann},
  title = {Phase diagram for {UPd}$_3$ subject to symmetry-breaking stress and magnetic field},
  journal = {Physical Review B},
  volume = {51},
  number = {17},
  pages = {11319--11324},
  year = {1995},
  doi = {10.1103/PhysRevB.51.11319},
  url = {https://doi.org/10.1103/PhysRevB.51.11319}
}

@article{Valiska2024PRM,
  author = {Vali\v{s}ka, M. and Haidamak, T. and Cabala, A. and Posp\'i\v{s}il, J. and Bastien, G. and Opletal, P. and Sakai, H. and Haga, Y. and Miyata, A. and Gorbunov, D. and Zherlitsyn, S. and Sechovsk\'y, V. and Prokle\v{s}ka, J.},
  title = {Dramatic elastic response at the critical end point in {UTe}$_2$},
  journal = {Physical Review Materials},
  volume = {8},
  pages = {094415},
  year = {2024},
  doi = {10.1103/PhysRevMaterials.8.094415},
  url = {https://doi.org/10.1103/PhysRevMaterials.8.094415}
}

@article{Tokiwa2023PRB,
  author = {Tokiwa, Y. and Sakai, H. and Kambe, S. and Opletal, P. and Yamamoto, E. and Kimata, M. and Awaji, S. and Sasaki, T. and Yanase, Y. and Haga, Y. and Tokunaga, Y.},
  title = {Anomalous vortex dynamics in the spin-triplet superconductor {UTe$_2$}},
  journal = {Physical Review B},
  volume = {108},
  number = {14},
  pages = {144502},
  year = {2023},
  doi = {10.1103/PhysRevB.108.144502},
  url = {https://doi.org/10.1103/PhysRevB.108.144502}
}

@article{Knebel2019,
  author  = {Knebel, Georg and Knafo, William and Pourret, Alexandre and Niu, Qun and Vali{\v{s}}ka, Michal and Braithwaite, Daniel and Lapertot, G{\'e}rard and Nardone, Marc and Zitouni, Abdelaziz and Mishra, Sanu and Sheikin, Ilya and Seyfarth, Gabriel and Brison, Jean-Pascal and Aoki, Dai and Flouquet, Jacques},
  title   = {Field-Reentrant Superconductivity Close to a Metamagnetic Transition in the Heavy-Fermion Superconductor {UTe$_2$}},
  journal = {Journal of the Physical Society of Japan},
  volume  = {88},
  number  = {6},
  pages   = {063707},
  year    = {2019},
  doi     = {10.7566/JPSJ.88.063707}
}

@article{Knebel2020,
  author  = {Knebel, Georg and Kimata, Motoi and Vali{\v{s}}ka, Michal and Honda, Fuminori and Li, DeXin and Braithwaite, Daniel and Lapertot, G{\'e}rard and Knafo, William and Pourret, Alexandre and Sato, Yoshiki J. and Shimizu, Yusei and Kihara, Takumi and Brison, Jean-Pascal and Flouquet, Jacques and Aoki, Dai},
  title   = {Anisotropy of the Upper Critical Field in the Heavy-Fermion Superconductor {UTe$_2$} under Pressure},
  journal = {Journal of the Physical Society of Japan},
  volume  = {89},
  number  = {5},
  pages   = {053707},
  year    = {2020},
  doi     = {10.7566/JPSJ.89.053707}
}

@article{Aoki2021,
  author  = {Aoki, Dai and Kimata, Motoi and Sato, Yoshiki J. and Knebel, Georg and Honda, Fuminori and Nakamura, Ai and Li, Dexin and Homma, Yoshiya and Shimizu, Yusei and Knafo, William and Braithwaite, Daniel and Vali{\v{s}}ka, Michal and Pourret, Alexandre and Brison, Jean-Pascal and Flouquet, Jacques},
  title   = {Field-Induced Superconductivity near the Superconducting Critical Pressure in {UTe$_2$}},
  journal = {Journal of the Physical Society of Japan},
  volume  = {90},
  number  = {7},
  pages   = {074705},
  year    = {2021},
  doi     = {10.7566/JPSJ.90.074705}
}

@article{Valiska2021,
  author  = {Vali{\v{s}}ka, M. and Knafo, W. and Knebel, G. and Lapertot, G. and Aoki, D. and Braithwaite, D.},
  title   = {Magnetic reshuffling and feedback on superconductivity in {UTe$_2$} under pressure},
  journal = {Physical Review B},
  volume  = {104},
  number  = {21},
  pages   = {214507},
  year    = {2021},
  doi     = {10.1103/PhysRevB.104.214507}
}

@article{Wu2025PNAS,
  author  = {Wu, Z. and Chen, H. and Weinberger, T. I. and Cabala, A. and Graf, D. E. and Skourski, Y. and Xie, W. and Ling, Y. and Zhu, Z. and Sechovsk{\'y}, V. and Vali{\v{s}}ka, M. and Grosche, F. M. and Eaton, A. G.},
  title   = {Superconducting critical temperature elevated by intense magnetic fields},
  journal = {Proc. Natl. Acad. Sci. U.S.A.},
  volume  = {122},
  number  = {2},
  pages   = {e2422156122},
  year    = {2025},
  doi     = {10.1073/pnas.2422156122}
}

@article{Wu2025PRL,
  author  = {Wu, Zheyu and Chen, Jiasheng and Weinberger, Theodore I. and Cabala, Andrej and Sechovsk{\'y}, Vladim{\'i}r and Vali{\v{s}}ka, Michal and Alireza, Patricia L. and Eaton, Alexander G. and Grosche, F. Malte},
  title   = {Magnetic Signatures of Pressure-Induced Multicomponent Superconductivity in {UTe$_2$}},
  journal = {Physical Review Letters},
  volume  = {134},
  number  = {23},
  pages   = {236501},
  year    = {2025},
  doi     = {10.1103/PhysRevLett.134.236501}
}

@article{Aoki2022dHvA,
  author  = {Aoki, Dai and Sakai, Hironori and Opletal, Petr and Tokiwa, Yoshifumi and Ishizuka, Jun and Yanase, Youichi and Harima, Hisatomo and Nakamura, Ai and Li, Dexin and Homma, Yoshiya and Shimizu, Yusei and Knebel, Georg and Flouquet, Jacques and Haga, Yoshinori},
  title   = {First Observation of the de Haas--van Alphen Effect and Fermi Surfaces in the Unconventional Superconductor {UTe$_2$}},
  journal = {Journal of the Physical Society of Japan},
  volume  = {91},
  number  = {8},
  pages   = {083704},
  year    = {2022},
  doi     = {10.7566/JPSJ.91.083704}
}

@misc{Vasina2026Calorimetry,
  author        = {Vasina, T. and Pfeiffer, M. and Borth, R. and Nicklas, M. and Amano Patino, M. and Lapertot, G. and Brison, J.-P. and Hassinger, E. and Knebel, G. and Braithwaite, D.},
  year          = {2026},
  archivePrefix = {arXiv},
  eprint        = {2603.29760},
  primaryClass  = {cond-mat.str-el}
}

@article{Avraham2001,
  author  = {Avraham, Nurit and Khaykovich, Boris and Myasoedov, Yuri and
             Rappaport, Michael and Shtrikman, Hadas and Feldman, Dima E. and
             Tamegai, Tsuyoshi and Kes, Peter H. and Li, Ming and
             Konczykowski, Marcin and van der Beek, Kees and Zeldov, Eli},
  title   = {Inverse melting of a vortex lattice},
  journal = {Nature},
  volume  = {411},
  pages   = {451--454},
  year    = {2001},
  doi     = {10.1038/35078021}
}

@article{Melnikovsky2026,
  author        = {Melnikovsky, L. A.},
  title         = {Superconducting Triple Point in {UTe$_2$}: Thermodynamics and Symmetry},
  journal       = {arXiv},
  year          = {2026},
  eprint        = {2606.05210},
  archivePrefix = {arXiv},
  primaryClass  = {cond-mat.supr-con},
  doi           = {10.48550/arXiv.2606.05210}
}

@article{Ishihara2023,
  author  = {Ishihara, Kota and Roppongi, Masaki and Kobayashi, Masayuki
             and Imamura, Kumpei and Mizukami, Yuta and Sakai, Hironori
             and Opletal, Petr and Tokiwa, Yoshifumi and Haga, Yoshinori
             and Hashimoto, Kenichiro and Shibauchi, Takasada},
  title   = {Chiral superconductivity in {UTe$_2$} probed by anisotropic low-energy excitations},
  journal = {Nature Communications},
  volume  = {14},
  pages   = {2966},
  year    = {2023},
  doi     = {10.1038/s41467-023-38688-y}
}

@misc{SupplementalMaterial,
  note = {See Supplemental Material appended to this manuscript for details of the sample preparation and measurement protocol, ultrasound modes and pulse-echo analysis, comparison of increasing- and decreasing-field sweeps, and temperature-dependent elastic measurements.}
}

\clearpage
\onecolumngrid

\setcounter{section}{0}
\setcounter{equation}{0}
\setcounter{figure}{0}
\setcounter{table}{0}
\renewcommand{\thesection}{S\arabic{section}}
\renewcommand{\theequation}{S\arabic{equation}}
\renewcommand{\thefigure}{S\arabic{figure}}
\renewcommand{\thetable}{S\arabic{table}}

\begin{center}
{\large\bfseries Supplemental Material for: Thermodynamic Evidence of Tetracritical Topology in the
$H$--$T$ Phase Diagram of UTe$_2$ for $H\parallel b$}\par
\vspace{0.8em}
Michal Vali\v{s}ka, Tetiana Haidamak, Andrej Cabala, Petr Proschek,
Andreas Hausprug, Sergei Zherlitsyn, and Vladim\'\i r Sechovsk\'y
\end{center}
\vspace{0.8em}

\section{Sample preparation and measurement protocol}

High-quality single crystals of UTe$_2$ were grown by the molten-salt-flux technique following previously reported procedures \cite{Sakai2022,Eaton2024}. Purified uranium metal and tellurium were reacted in an equimolar NaCl--KCl flux using a carbon crucible sealed in an evacuated quartz ampoule. The uranium starting material was refined by solid-state electrotransport before use, and the starting materials were handled under an argon atmosphere to minimize oxidation. After growth, the salt flux was dissolved in water and bar-shaped single crystals were mechanically separated, rinsed, and stored under an argon atmosphere prior to characterization and ultrasound measurements.

The crystal used for the ultrasound measurements showed a superconducting
transition with $T_c>2$~K. Its crystallographic orientation was determined
by Laue diffraction. The selected crystal was naturally bar-shaped and
therefore required no cutting or polishing, minimizing the possibility of
introducing additional strain through mechanical processing. Its dimensions
were approximately $4.8\times0.7\times0.7$~mm$^3$ along the crystallographic
$a$, $b$, and $c$ axes, respectively. The long axis of the crystal was
parallel to $a$, while the ultrasound propagation direction was parallel
to $c$ for all measured modes.

LiNbO$_3$ piezoelectric transducers were bonded to the sample using a Thiokol adhesive. Measurements were performed at operating frequencies between 34 and 140~MHz, in magnetic fields up to 18~T and at temperatures down to 0.33~K. The magnetic field was applied along the crystallographic $b$ axis and swept at a rate of 0.5~T~min$^{-1}$.

For each field-sweep sequence at a given temperature, the sample was first warmed above the superconducting transition, maintained there for approximately 10~min, and subsequently cooled in zero field to the target temperature. Increasing- and decreasing-field sweeps were then recorded at the same sweep rate. This zero-field-cooling protocol was repeated before measurements at each temperature.

Complementary transverse magnetostriction $\Delta l_c/l_c$ for $H\parallel b$ was measured using a high-resolution capacitive dilatometer on a crystal from the same growth batch. The dilatometer measured the relative length change along the crystallographic $c$ axis, thereby probing the equilibrium $c$-axis strain and providing a static counterpart to the dynamic $C_{33}$ elastic response.

\section{Ultrasound modes and pulse-echo analysis}

For an acoustic mode, the elastic constant is related to the sound velocity by
\begin{equation}
C_{ij}=\rho v^{\,2},
\label{eqS:Cij_velocity}
\end{equation}
where $\rho$ is the mass density and $v$ is the sound velocity of the corresponding acoustic mode \cite{Luthi2005}. In an ultrasound experiment, the measured elastic mode is specified by the propagation vector $\mathbf{q}$ and the polarization $\mathbf{u}$ of the acoustic wave. In the present work all measured modes propagate along the $c$ axis: the longitudinal geometry $\mathbf{q}\parallel c$, $\mathbf{u}\parallel c$ probes $C_{33}$, while the transverse geometries $\mathbf{q}\parallel c$, $\mathbf{u}\parallel b$ and $\mathbf{q}\parallel c$, $\mathbf{u}\parallel a$ probe $C_{44}$ and $C_{55}$, respectively.

Magnetostriction provides a complementary probe of the strain-dependent thermodynamic potential $f$. In the absence of external stress, the equilibrium strain is determined by the minimum condition
\begin{equation}
\left(\frac{\partial f}{\partial \varepsilon_i}\right)_{T,H}=0,
\label{eqS:minimum_condition}
\end{equation}
so that the measured length change $\Delta l_i/l_i \equiv \varepsilon_i(H)$ corresponds to the field evolution of the equilibrium strain.

In the pulse-echo experiment, the directly measured quantity is the relative frequency shift $\Delta f/f$. For small changes,
\begin{equation}
\frac{\Delta v}{v}=
\frac{\Delta f}{f}+\frac{\Delta l}{l}-\frac{\Delta\varphi}{\varphi}.
\label{eqS:dv_over_v}
\end{equation}
Here $\varphi$ is the phase of the acoustic wave, which is kept constant in our ultrasound experiments, so that $\Delta\varphi/\varphi = 0$. The magnetostriction term in Eq.~(\ref{eqS:dv_over_v}) is only of order $10^{-6}$, two orders of magnitude smaller than $\Delta f/f$, and is therefore neglected without affecting the anomaly positions.

\section{Comparison of increasing- and decreasing-field sweeps}

To assess possible hysteresis associated with the high-field anomaly, we compare increasing- and decreasing-field $C_{33}(H)$ sweeps between 0.33 and 1.1~K, where the anomaly associated with the internal superconducting boundary is clearly resolved. Upon increasing field, this anomaly marks the entry into the region denoted SC1+SC2 in the phase diagram, while the decreasing-field sweep crosses the same boundary in the reverse direction.

\begin{figure}[p!]
    \centering
    \includegraphics[
        width=0.88\linewidth,
        height=0.62\textheight,
        keepaspectratio
    ]{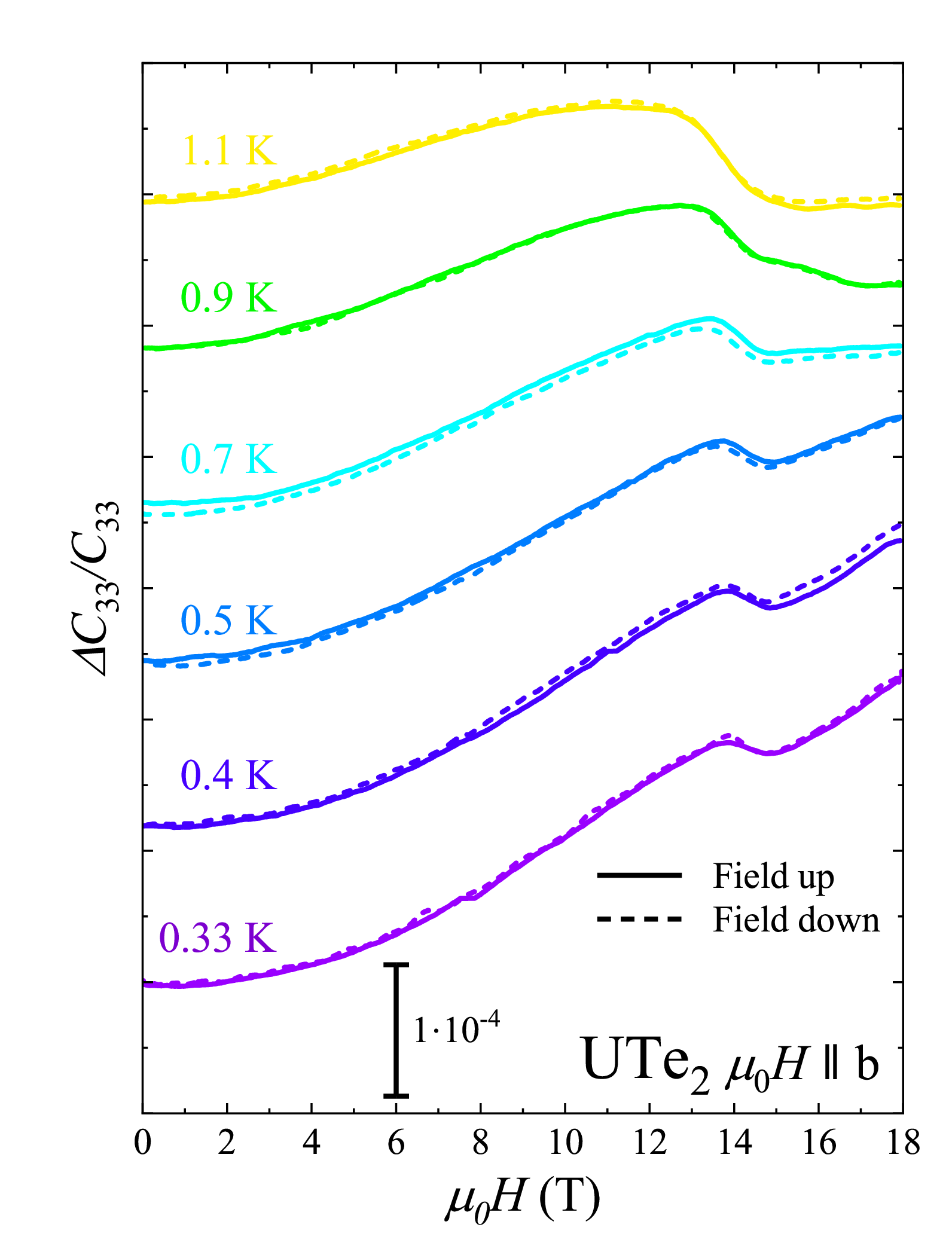}
    \caption{
    Comparison of increasing- and decreasing-field $C_{33}(H)$ sweeps
    measured between 0.33 and 1.1~K.
    At each temperature, the traces were aligned using their respective
    low-field values and shifted together by a common vertical offset.
    The internal-boundary anomaly occurs at the same field for both sweep
    directions within the experimental resolution, with no measurable
    hysteresis of its position.
    }
    \label{figS:hysteresis}
\end{figure}

For each temperature, the increasing- and decreasing-field traces were referenced to their respective low-field values by subtracting the first and last data point, respectively. The resulting pair of curves was then shifted by a common vertical offset for clarity. As shown in Fig.~\ref{figS:hysteresis}, the anomaly occurs at the same field within the experimental resolution for both sweep directions, and no systematic hysteresis of its position is resolved over the investigated temperature range.

The observed reversibility strongly disfavors irreversible vortex pinning or depinning as the origin of the anomaly. It does not, by itself, rigorously exclude a reversible equilibrium vortex-matter transition, as discussed in the main text.

\section{Temperature-dependent elastic response}

\begin{figure}[p!]
    \centering
    \includegraphics[
        width=0.86\linewidth,
        height=0.58\textheight,
        keepaspectratio
    ]{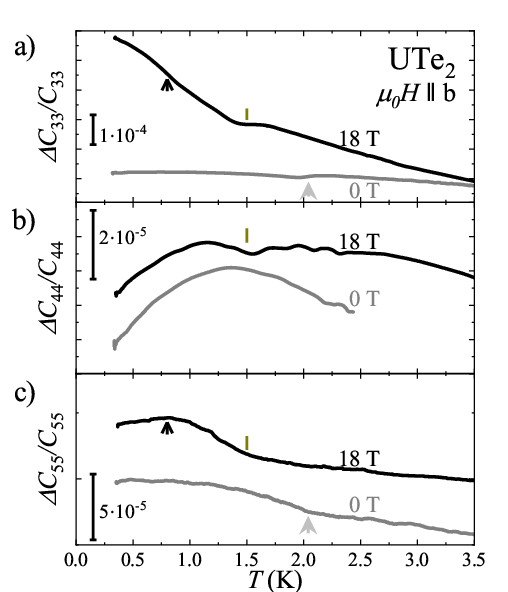}
    \caption{
    Temperature dependence of the elastic response at selected magnetic
    fields.
    (a) $C_{33}(T)$ for $\mathbf{q}\parallel c$ and
    $\mathbf{u}\parallel c$.
    (b) $C_{44}(T)$ for $\mathbf{q}\parallel c$ and
    $\mathbf{u}\parallel b$.
    (c) $C_{55}(T)$ for $\mathbf{q}\parallel c$ and
    $\mathbf{u}\parallel a$.
    Each panel includes the 0 and 18~T curves. The marked anomalies denote
    transition points used to construct the phase diagram in Fig.~2 of the
    main text. Vertical scales are indicated by the scale bars.
    }
    \label{figS:temperature_sweeps}
\end{figure}

Figure~\ref{figS:temperature_sweeps} summarizes temperature sweeps of $C_{33}$, $C_{44}$, and $C_{55}$ at selected magnetic fields. The marked anomalies provide complementary transition points used to construct the phase diagram shown in Fig.~2 of the main text.

\end{document}